\documentclass[referee]{aa}
\usepackage{graphicx}
\usepackage{hyperref}
\usepackage{amsmath}
\usepackage[varg]{txfonts}
\usepackage{color}
\usepackage[english]{babel}

\newcommand{\cm}{cm$^{-1}$}

\newcommand{\2}{$_{2}$}

\newcommand{\4}{$_{4}$}

\def\a0{{$a_{\rm 0}$}}

\titlerunning{Hybrid line list for CH$_4$}
\authorrunning{Yurchenko et al.}

\begin{document}

\title{A hybrid line list for CH$_4$ and hot methane continuum }

%
\author{Sergei N. Yurchenko \thanks{The corresponding author: Sergei N. Yurchenko; E-mail: s.yurchenko@ucl.ac.uk} \inst{1} \and David~S.~Amundsen\inst{2,3,4} \and Jonathan Tennyson \inst{1} \and Ingo P. Waldmann \inst{1}}

\institute{%
Department of Physics and Astronomy, University College London, London WC1E 6BT, United Kingdom
\and
Astrophysics Group, University of Exeter, Exeter, EX4 4QL, United Kingdom
\and
Department of Applied Physics and Applied Mathematics, Columbia University, New York, NY 10025, USA
\and
NASA Goddard Institute for Space Studies, New York, NY 10025, USA}


\date{\today}

\abstract{}{ Molecular line lists (a catalogue of transition
  frequencies and line strengths) are important for modelling
  absorption and emission processes in atmospheres of different
  astronomical objects, such as cool stars and exoplanets. In order to
  be applicable for high temperatures, line lists for molecules like methane
  must contain billions of transitions, which makes their direct
  (line-by-line usage) application in radiative transfer calculations
  impracticable. Here we suggest a new, hybrid line list format to mitigate this problem,  based on the idea of temperature-dependent absorption continuum.}
{The line list is  partitioned into a large set of relatively weak lines and a small  set of important, stronger lines. The weaker lines are then used either to  construct a temperature-dependent (but pressure-independent) set of  intensity cross sections or are blended into a greatly reduced set of `super'-lines. The strong lines are kept in the form of   temperature independent Einstein~$A$ coefficients.
  }
{  A line list for methane (CH$_4$) is constructed
  as a combination of 17 million strong absorption lines relative to
  the reference absorption spectra and a background methane continuum
  in two temperature-dependent forms,  of cross sections  and
  super-lines. This approach eases the use of large high temperature
  line lists significantly as the computationally expensive
  calculation of pressure dependent profiles (e.g. Voigt) only need to be performed for a relatively small number of lines.
  Both the line list and cross sections were generated using a new 34
  billion methane line list (34to10), which extends the 10to10 line
  list to higher temperatures (up to 2000~K). The new hybrid scheme can be
  applied to any large line lists containing billions of transitions. We recommend to use super-lines generated on a high resolution grid based on resolving power ($R =$ 1,000,000) to model the molecular continuum as a more flexible alternative to the temperature dependent cross sections.}
{}


\maketitle

\titlerunning{A hybrid line list for CH$_4$ }
\authorrunning{S.N. Yurchenko et. al.}

\section{Introduction}

Methane is one of the key absorbers in the atmospheres of exoplanets
and cool stars. Due to a large number of relatively strong lines (up to
several billion) at high temperatures, the calculation of cross
sections becomes extremely computationally expensive. The contribution
of each line to the total absorption must be taken into account by
summing their individual cross sections, usually computed using Voigt
profiles, on a grid of wavelengths. To make radiative transfer
calculations using these line lists more tractable the line lists are
usually converted into pre-computed tables of temperature and pressure
dependent cross sections, or $k$-coefficients, for specific
atmospheric conditions (temperature, pressure,
broadeners)~\citep{14AmBaTr.broad,HELIOS}. Subsequent radiative
transfer calculations interpolate in these tables. However, the calculation of these cross sections and $k$-coefficients still require
the contributions of all lines to be summed, if only once for each
atmospheric condition. Both pre-tabulated cross sections and
$k$-coefficients are less flexible than a line-by-line approach, but
computationally more efficient.

As part of the ExoMol project \citep{jt528} we have produced an
extensive line list for methane ($^{12}$CH$_4$), called 10to10
\citep{jt528}, containing almost 10 billion transitions. The
line list was constructed to describe the opacity of methane for
temperatures up to 1500~K. The 10to10 line list been has shown to be
important for modelling the atmospheres of brown dwarfs and exoplanets
\citep{jt572,15CaLuYu.dwarfs,16AmMaBa}, and has been used as an
input in a number of models such as TauREX \citet{jt593,jt611},
NEMESIS \citep{NEMESIS}, VSTAR \citep{12BaKe,jt572},
ATMO~\citep{15TrAmMo,16TrAmCh,16DrTrBa} and the UK Met
Office global circulation model (GCM) when applied to hot
Jupiters~\citep{16AmMaBa}. The ExoMol database contains line lists for
about 40 other molecular species and has recently been upgraded
\citep{jt631}. The line lists for polyatomic molecules usually contain
more than 10 billion lines; examples include phosphine (PH$_3$)
\citep{jt592}, hydrogen peroxide (H$_2$O$_2$)
\citep{jt620}, formaldehyde (H$_2$CO) \citep{jt597} and
SO$_3$ \citep{jt641} \citep[see also our review of molecular line lists][]{jt693}.

A promising alternative to the line-by-line approach was recently
proposed by \citet{15HaBeBa.CH4}, where an accurate experimental line
list of the strongest CH$_4$ transitions was complemented by a set of
experimental quasi-continuum cross sections, measured for a set of
different temperatures.  \citet{TheoReTS} recently proposed an
alternative, super-line (SL), approach to speed up the line-by-line
calculations. The idea is to build intensity histograms from
transition intensities binned for a given temperature into wavenumber
grid points. Each wavenumber bin is then treated as a super-line for
computing cross sections for different line profiles,
which brings the computational cost of a line-by-line
approach almost down to that using pre-tabulated cross sections. The
serious disadvantage, however, is that only very simplistic line
profiles, ones which do not depend on quantum numbers, can be used.
Indeed, each super-line loses memory of its upper and lower states,
only the wavenumber is preserved. This is not a problem for the
Doppler profile as it does not depend on quantum numbers. However,
pressure-dependent profiles such as Voigt profiles often show strong dependence
on the rotational $J$ and other quantum numbers, which cannot be
modelled using the SL approach.

In the present work we combine these two approaches and provide a
synthetic `hybrid' line list for methane using the following
compilation of data: (i) a line list of `strong' $N_{\rm str}$ lines
given explicitly using the ExoMol format \citep{jt542,jt631}; (ii) all
other $N_{\rm weak}$ `weak' lines are converted into
a temperature-dependent, but pressure independent background continuum.
Thus the aim of this work is to select the most important lines (both
the strongest and sensitive to the variation of line profiles with
pressure and broadener) for the direct line-by-line treatment, while
the rest are processed either as cross sections or as
super-lines \citep{TheoReTS}. 
The hybrid approach has the potential to retain the key features of
line lists and to significantly ease the computation of total cross
sections and $k$-coefficient tables (including both weak and strong
lines). We investigate two approaches to represent the
temperature-dependent continuum: (i) using pressure independent
cross sections described by the Doppler profile and (ii) using the
profile-free histograms (super-lines).

As demonstrated by \citet{14ReNiTy.CH4} and \citet{17NiReTy}, in order to extend the
temperature coverage of the 10to10 CH$_4$ line list the lower state energy
threshold should be increased compared that used by \citet{jt564}. Our
10to10 line list was based on the lower state energy threshold $E_{\rm
  max}^{\rm lower}=8000$~\cm\ which was estimated to be sufficient for
temperatures up to 1500~K. In this work we  extend the 10to10
line list by increasing $E_{\rm max}^{\rm lower}$ to 10,000~\cm, which
should extend the temperature coverage to about 2000~K. To be
consistent with the extension of the lower state energy threshold, the
rotational coverage had to be increased from $J_{\rm max} = 46$ used by
\citet{jt564} to about $J_{\rm max} = 50$. The cost of this
improvement, however, is a dramatic increase of the number of lines,
from 9.8~billion to 34~billion. The resulting `34to10' line list is
used in this work to build a continuum absorption model for methane as
described above.

The partitioning of the 34~billion line list into a set of $N_{\rm
  str}$ `strong' lines and $N_{\rm weak}$ `weak' lines is presented in
Section~\ref{s:s-w}, where we also define and test the `strong'/`weak'
partitioning. In Section~\ref{s:cross} our continuum model is tested by comparing
it to the traditional approach of explicitly summing up the cross
section contributions from all lines, at different temperature and pressure.
Section~\ref{s:result} presents our final results.


\section{Strong/Weak line list partitioning}
\label{s:s-w}

In the following, the new line list for methane, which we have named
34to10, is used in all our examples. The line list is an extension of
the 10to10 line list, produced using the same computational approach
\citep{jt564} by extending the lower state energy range from
8,000~\cm\ to $10,000$~\cm. Calculations were performed with nuclear
motion code TROVE \citep{TROVE}. As before, here calculations used a
spectroscopically-determined potential energy surface \citep{jt564}
and {\it ab initio} dipole moment surfaces \citep{jt555}.
The new line list contains 8,194,057
energies below 18,000~\cm\ and 34,170,582,862 transitions covering
rotational excitations up to $J_{\rm max} = 50$. The calculation of
the additional 28 billion transitions took approximately 5 million
CPU hours on the Cambridge High Performance Computing Cluster
Darwin. The wavenumber coverage, however, is kept the same as in the
10to10 line list, which means that the region from 10,000 to
12,0000~\cm\ is less complete for the target temperature of 2000~K.
All other computational components (potential energy and dipole moment
surfaces, basis sets etc.) are the same as in \citep{jt564}.


In order to mitigate the difficulty of using such an extremely large line list, we
propose dividing it into two sub-sets, responsible for strong and weak absorptions.
The first question is how to define and separate `strong' and `weak' transitions.
Due to the large dynamic variation of the methane intensities, a
single intensity threshold would be not optimal due to the following factors:
(i) In regions of very strong
bands many lines with moderate intensities are barely visible, while
weak lines which lie between the main bands can be relatively
important. (ii) The definition of `strong' and `weak' must be
temperature dependent as `hot' bands,
which are weak at low temperatures due to the Boltzmann factor, become
stronger with increasing population of excited lower states at higher
temperatures. (iii) At the same time the intensities of the
fundamentals and overtones decrease with temperature due to the
decrease of their relative population (e.g. due to larger partition
function). (iv) Finally, even relatively weak lines at longer
wavelengths are very sensitive to pressure variations due to their
relatively lower density. It is therefore necessary to take these factors into
account when defining the intensity partitioning thresholds.

\begin{figure}[!htbp]
  \includegraphics[width=0.48\textwidth]{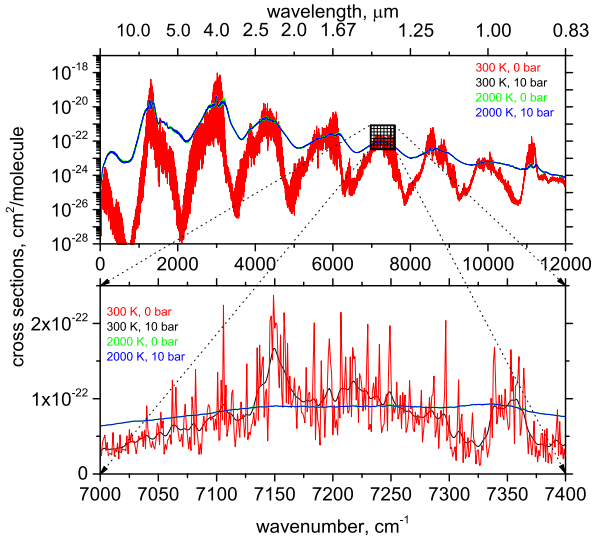}
\caption{\label{f:cross:ref} Reference cross sections obtained using the Doppler profile at $T=300$~K and $T=2000$~K on the uniform $\Delta \tilde{\nu} = 10$ wavenumber grid. The green line ($T=2000$~K and $P=0$~bar) is almost identical to the blue line ($T=2000$~K and $P=50$~bar) at this region and for this scale and thus can be barely seen.}
\end{figure}

To aid building a `strong'/`weak' partitioning, we introduced a reference CH\4\ opacity $\alpha_{\rm ref} (\tilde{\nu})$ based on two temperatures, $T_1 = 300$~K and $T_2=2000$~K, and two pressures, $P_1=0$~bar and $P_2=50$~bar, on a wavenumber grid of $\Delta \tilde{\nu}  = 1$ \cm\ ($\tilde{\nu} =  0 \ldots 12000$~\cm) by choosing the maximum cross section value amoung these four at each wavenumber grid point $k$:
\begin{equation}\label{e:Imax}
  \alpha_{\rm ref}(\tilde{\nu}_{k}) = {\rm max} (\alpha_{300}^{P=0},\alpha_{2000}^{P=0},\alpha_{300}^{P=50},\alpha_{2000}^{P=50}).
\end{equation}
The reference average intensities (cm/molecule) can then be defined as:
\begin{equation}\label{e:ref}
\bar{I}(\tilde{\nu}_k)  \equiv  \alpha_{\rm ref}(\tilde{\nu}_{k}) \Delta \tilde{\nu} .
\end{equation}
Figure~\ref{f:cross:ref} shows the
reference cross section curve used here for the 34to10 line list.

\begin{figure}[!htbp]
  \includegraphics[width=0.48\textwidth]{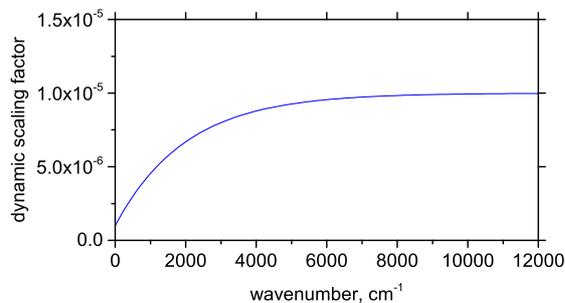}
\caption{\label{f:scaling}
Dynamic scaling factor used in Eq.~\protect\ref{e:scale}. }
\end{figure}

We then define the `strong'/`weak' partitioning using two criteria, one dynamic and one static.
\textit{Static}: All lines stronger than the threshold $I_{\rm thr}$ are automatically taken into the `strong' section (e.g. $I_{\rm thr} = 10^{-25}$ cm/molecule). \textit{Dynamic}:
The line $\tilde{\nu}_{fi}$ from the wavenumber bin $k$ ($\tilde{\nu}_{fi} \in [\tilde{\nu}_{k}-0.5$ \cm, $\tilde{\nu}_{k}+0.5$ \cm$] $) is `strong' if all four reference absorption intensities are stronger than the reference (average) $\bar{I}_{\tilde{\nu}_k}$ intensity by some scaling factor $C_{\rm scale}$ (e.g. stronger than $10^{-5} \times \bar{I}_{\tilde{\nu}_{k}}$). The scaling factor $C_{\rm scale}$ is made wavenumber dependent using the following exponential form, also shown in Fig.~\ref{f:scaling}:
\begin{equation}\label{e:scale}
  C_{\rm scale} (\tilde\nu) =  10^{-5} \times \left[1-0.9\, e^{-0.0005\tilde\nu}\right].
\end{equation}
This scaling is necessary to take into account the importance of the varying density of lines at different spectroscopic regions for the accurate description of the line profiles: the smaller number of lines at the longer wavelengths means the cross sections are more sensitive to the shape of the profiles as well as to the sampling of the grid points. At the shorter wavelengths the spectrum is smoothed out by the large number of overlapping lines, which is therefore less sensitive to these factors. With this expression we thus assume a quasi-exponential increase of the density of lines vs wavenumber, or, colloquially, a quasi-exponential decrease of their importance.

\begin{figure}[!htbp]
  \includegraphics[width=0.48\textwidth]{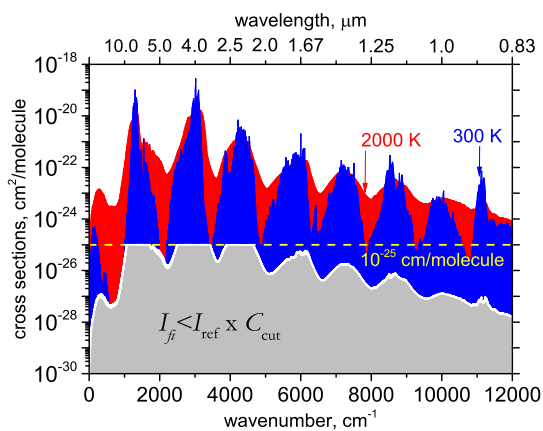}
\caption{\label{f:strong:example}
An example of the intensity partitioning for $I_{\rm thr} = 10^{-25}$~cm/molecule and $C_{\rm scale} = 10^{-5} $. The dashed line indicates the $I_{\rm thr}$ threshold; the blue ($T=300$~K) and red ($T=2000$~K) areas are the regions of the `strong' lines; the grey area at the bottom indicates all transitions which were excluded from the line list to form the `weak' lines of the continuum. Here all cross sections were obtained using the Doppler profile on a grid of 10~\cm. }
\end{figure}

\begin{figure}[!htbp]
  \includegraphics[width=0.48\textwidth]{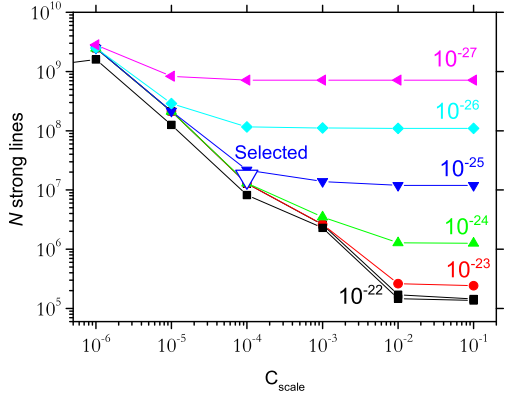}
  \caption{\label{f:N:lines} Number of `strong' lines for different partitionings.  }
\end{figure}

Figure~\ref{f:strong:example} and Figure~\ref{f:N:lines}  illustrate how these partitioning criteria affect the absorption cross sections and size of the strong and weak lines partitions, respectively,  using the constant scale factor $C_{\rm scale}$ for simplicity. For example, the combination ($C_{\rm scale} = 10^{-2}$, $I_{\rm thresh} \times$ [cm$/$molecules]$^{-1}$  = $10^{-23}$), with $C_{\rm scale}$ constant,  leads to 262,470 lines.  Using the scale factor $C_{\rm scale} = 10^{-5}$ increases the number of strong lines by an order of magnitude. For example, for the partitioning ($10^{-5}$,$10^{-21}$) we obtain 125 million strong lines.
The dynamic partitioning defined by Eq.~\ref{e:scale} in combination with $I_{\rm thresh} $  = $10^{-23}$~cm$/$molecules is also shown in Fig.~(\ref{f:N:lines}) as a large triangle. This partitioning is also our preferred choice used in the following discussions as well as to construct the hybrid line list presented in this work. It results in 17 million selected lines (16,776,857) as part of the strong section, out of the original $34 \times 10^{10}$. This is a significant reduction and should ease line-by-line calculations significantly. The remaining lines are  converted into temperature-dependent histograms (super-lines) and/or cross sections to form our methane quasi-continuum, which is described below. By comparison, the HITRAN~2012~\citep{HITRAN2012} databases contains 336,830 $^{12}$CH\4\ transitions.

\section{Absorption continuum cross sections}
\label{s:cross}

\subsection{Quasi-continuum from the Doppler line profile}

The main difficulty associated with modelling cross sections (i.e. dressing lines with appropriate absorption profiles) is the pressure effect, requiring line shapes to be described using Lorentzian profiles (high pressure), Voigt profiles (moderate to high pressure) or even more sophisticated profiles \citep{jt584}. The Doppler profile (zero pressure), however, is much simpler: it is fast to compute, with a simple parametrisation of the line width (mass and frequency dependent only), an no dependence on the transition quantum numbers, mixing ratios of broadeners etc. \citep{14AmBaTr.broad}.

\begin{figure}[!htbp]
  \includegraphics[width=0.48\textwidth]{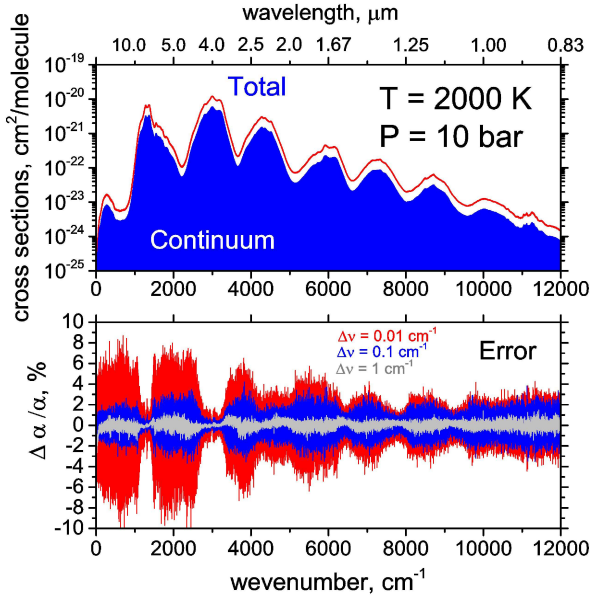}
\caption{\label{f:0-12000}
Upper display: Methane continuum  at 2000~K, $P=10$~bar (blue) and the total absorption (red).
The lower display shows the relative differences of the $P=0$ and $P=10$~bar continuum cross sections for the three wavenumber grids of $\Delta \tilde{\nu} = 0.01$~\cm\ (red), 0.1~\cm\ (blue) and 1~\cm\ (grey).}
\end{figure}

We will assume that the `weak'-lines quasi-continuum forms a nearly featureless background that is not very sensitive to the variation of pressure (at least for moderate pressures). This means the exact shape of the lines that form this quasi-continuum is relatively unimportant and can be modelled using a pressure-independent, temperature-dependent profile. Basically, our assumption is that any realistic line profile would be applicable as long as it preserves the area as the frequency integrated cross section of each line.
 In order to illustrate this approach, we show in Figure~\ref{f:0-12000} the quasi-continuum cross section from the `weak' lines. The cross section was computed at 2000~K using the ExoCross code \citep{ExoCross} as described by \citet{jt542} for our selected partitioning using a Doppler line profile.


In order to benchmark the zero pressure Doppler-based model of the continuum absorption we also computed the corresponding cross sections using the Voigt line profile at $P=10$~bar, $T=2000$~K. We use the simple ExoMol pressure-broadening diet of \citet{jt684} to describe the Voigt broadening of CH\4\ lines by 100~\%\ H\2. The $J$ dependence of the pressure-broadened half-width, $\gamma$, is similar to that used by \citet{14AmBaTr.broad}, and the temperature-dependence exponent, $n$, is assumed to be a constant. The broadening model is provided as part of the supplementary material to this paper.
A grid spacing of $\Delta \tilde{\nu} = 0.01$~\cm\ was chosen.
Figure~\ref{f:0-12000} (bottom display) also shows the relative difference between the Doppler-based continuum ($P=0$) and the realistic $P=10$~bar continuum (Voigt)  on three grids of 0.01, 0.1 and 1~\cm\  at $T=2000$~K. The grid of 0.01~\cm\ shows the fluctuations of the error
within 2-8~\%. Here the relative difference of cross sections is defined as follows
\begin{equation}\label{e:error}
\frac{\Delta \alpha(\tilde\nu)}{\alpha(\tilde\nu)} = \frac{\alpha(\tilde\nu)_{P}- \alpha_{P=0}(\tilde\nu)}{\alpha_{P}^{\rm tot}(\tilde\nu)},
\end{equation}
where $\alpha_{P=0}$,  $\alpha_{P} $ and $\alpha_{P}^{\rm Tot}$ are the $P=0$ (Doppler) continuum, $P\not = 0 $ continuum (Voigt) and the $P\ne 0$ total cross section, respectively. The largest error is for the long wavelength region, characterized by the weakest intensities and least densities of lines. In this region the Doppler-broadened lines become increasingly narrow, which makes the cross section to be very sensitive to the grid sampling used.  The best agreement is in the spectral regions with large cross sections and at short wavelengths, where the density is highest. Using coarser grids of 0.1 or 1~\cm\ drops the fluctuations to within 4 and 1.5~\%, respectively. The total integrated difference should be zero by definition since the area of the Voigt profile is conserved (subject to the numerical error). However, we note that unless the background lines are optically thin the resulting integrated flux will not be conserved.

\begin{figure*}
  \includegraphics[width=0.5\textwidth]{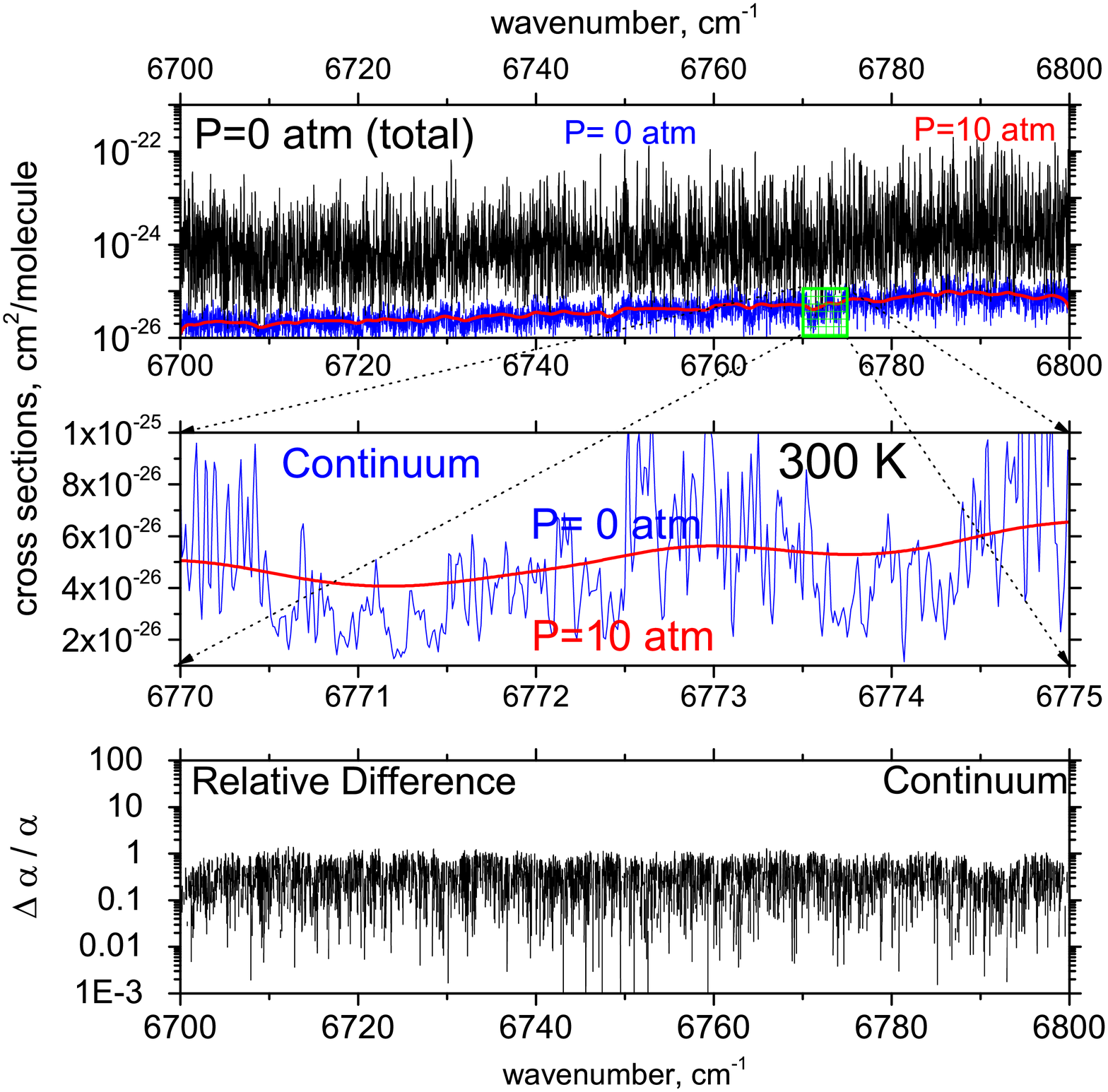}
  \includegraphics[width=0.5\textwidth]{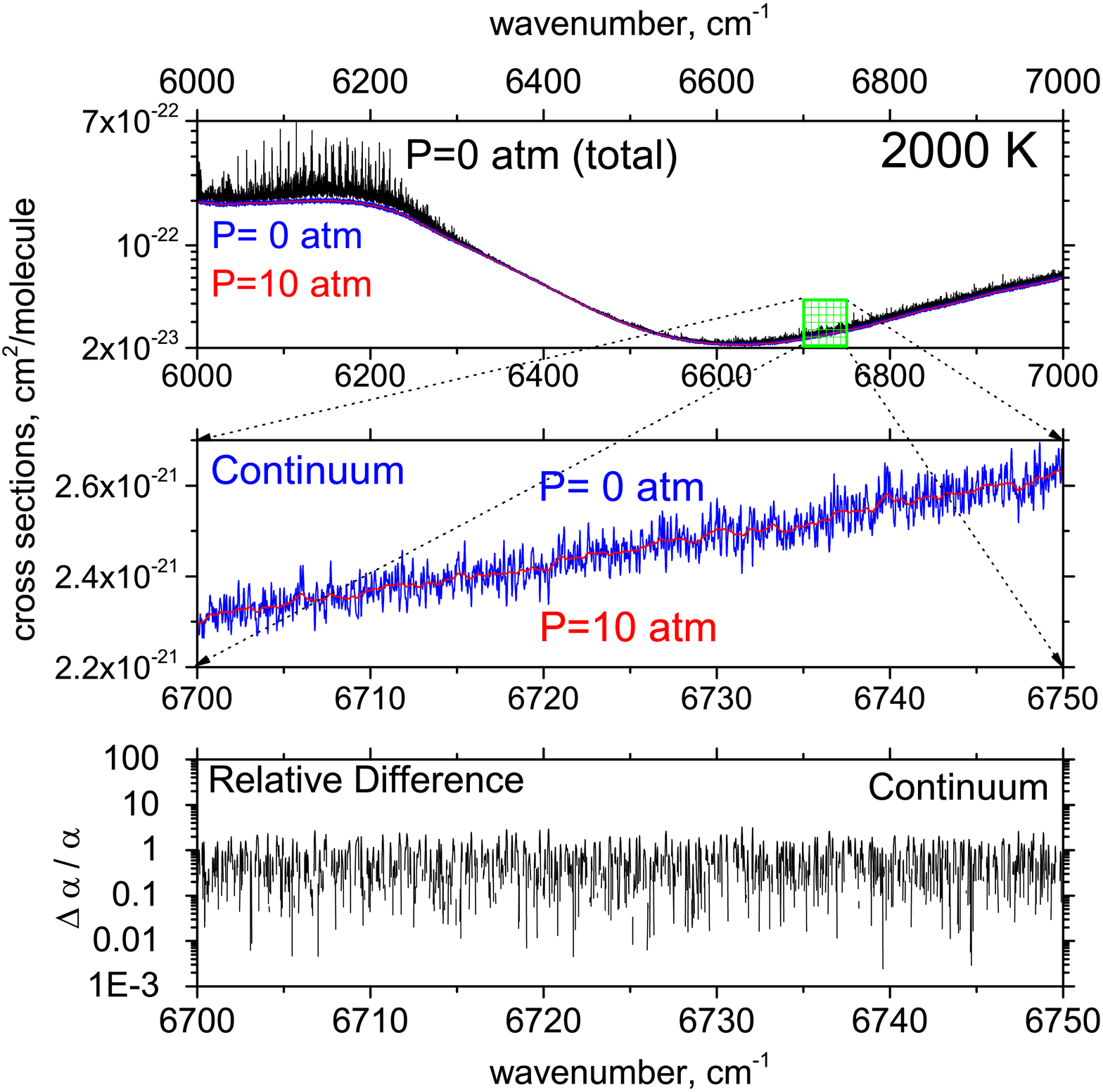}
\caption{\label{f:6000}
Comparison of the $P=0$ and $P=10$~bar cross sections for 300~K (left) and 2000~K (right): black (total $P=0$), blue (continuum $P=0$) and red (continuum  $P=10$). The middle panels are a blow up of the continuum, also for $P=0$ and $P=10$~bar, which are almost indistinguishable on the upper panels. The lower row shows the relative difference between the $P=0$ and $P=10$~bar continuum cross sections as defined in Eq.~(\protect\ref{e:error}). The integrated area of the relative difference is  0.06~\%\ over the region 6700 -- 6750~\cm.  A wavenumber grid of $\Delta \tilde{\nu} = 0.01$~\cm\ was used. }
\end{figure*}

A more detailed example of the $P=0$ and $P=10$~bar cross sections for the region 6000 -- 7000 ~ \cm\ is shown in Fig.~\ref{f:6000} for 300~K (left) and 2000~K (right). Even on the very small scale (see a zoom-in in the middle panels of this figure) the $P=0$ and $P=10$~bar continuum cross sections are almost identical: the difference between the two continuum curves ($P=0$ and $P=10$~bar) is barely seen. The bottom panels of Fig.~\ref{f:6000} show absolute relative differences $|\Delta \alpha(\tilde\nu)|/\alpha(\tilde\nu)$ between these two cross sections. For our partitioning a 1-2~\%\ accuracy (measured as the relative difference between these two profiles) is achieved for this region. In fact, the difference is not systematic, therefore the integrated effect should be even smaller. For example, integration of the relative difference $\Delta \alpha(\tilde\nu)$ for $T=$~2000~K in the region 6700 -- 6800~\cm\ gives an error of only 0.004~\%\ using the grid spacing of $\Delta \tilde{\nu} = 0.01$~\cm. The fluctuations for $T = 300$~K between the high and zero pressure cases are slightly higher, but still within approximately 1-2~\%. The integrated relative difference in this case is about 0.06~\%\  (6700 -- 6750~\cm, see Fig.~\ref{f:6000}).

\begin{figure}[!htbp]
  \includegraphics[width=0.48\textwidth]{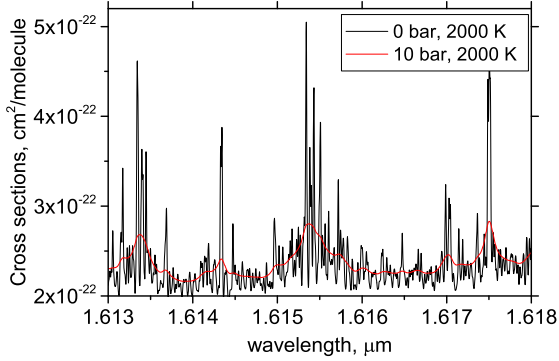}
\caption{\label{f:P0vsP10}
Comparison of the $P=0$ and $P=10$~bar line profiles used to generate cross sections at 2000~K in the region of 1.615~$\mu$m. The $a_0$ model with the $J$-independent line broadening was used. The wavenumber grid  $\Delta \tilde{\nu}$ is $0.01$~\cm.  }
\end{figure}

The corresponding line shapes are very different at these temperatures and pressures. The total $P=0$ and $P=10$~bar cross sections  have very different profiles as also illustrated in Fig.~\ref{f:P0vsP10}. However the difference
between continuum curves is negligible (see also Fig.~\ref{f:6000}).

\subsection{Super-line approach}

In this section we consider temperature-dependent lists of super-lines \citep{TheoReTS}, which present a more flexible alternative to the Doppler-broadened continuum in terms of the line-profile modelling. The super-lines are constructed as temperature-dependent intensity histograms as follows
\citep[see also detailed instruction in][]{TheoReTS}. The wavenumber range $[\tilde\nu_{A},\tilde\nu_{B}]$ is divided into $N$ frequency bins, each centered around a grid point $\tilde\nu_k$. Here we assume a general case of non-equidistant grids with variable widths $\Delta \tilde{\nu}_k$. For each $\tilde\nu_k$ the total absorption intensity ${I}_k(T)$ is computed as a sum of absorption line intensities $I_{if}$
\begin{equation}
\label{eq:int}
I_{if} = \frac{1}{8\pi c \tilde{\nu}^2} \frac{g_{\rm ns}(2J' + 1)}{Q(T)} A_{if} \exp\left( \frac{-c_2 \tilde{E}''}{T} \right) \left[1-\exp\left({\frac{c_2\tilde{\nu}}{T}}\right)\right] ,
\end{equation}
from all $i\to f$ transitions   falling into the wavenumber bin $[\tilde\nu_k-\Delta \tilde{\nu}_k/2 \ldots \tilde\nu_k+\Delta \tilde{\nu}_k/2 ] $ at the given temperature $T$. Here $A_{if}$ is the Einstein A coefficient ($s^{-1}$),  $c$ is the speed
of light (cm~$s^{-1}$), $Q(T)$ is the partition function, $\tilde{E}''$ is the lower state term value (\cm), $c_2$ is the second radiation constant (K~cm), $g_{\rm ns}$ is the nuclear statistical weight, $J'$ is the rotational angular momentum quantum number of the upper state and
$I_{if}$ is the line intensity or absorption coefficient (cm$^2$/molecule \cm).
Each grid point $\tilde\nu_k$ is then treated as a line position of an artificial transition   (super-line) with an effective absorption intensity ${I}_k(T)$. The `super'-line lists  are then formed as catalogues of these artificial transitions $\{\tilde\nu_k, {I}_k(T)\}$ with pre-computed intensities ${I}_{k}$. This can be compared to the temperature-independent ExoMol-type $\{\tilde\nu_{if}, A_{if}\}$  or temperature-dependent HITRAN-type $\{\tilde\nu_{if}, I_{if}(T)\}$ line lists.

As in the case of the conventional line lists, the super-lines can be used in line-by-line modelling of absorption cross sections, which significantly reduces the  computational costs. Indeed, each super-line
can be dressed with the corresponding line profile to generate actual cross sections for the corresponding $T$ and any given pressure broadening,  providing that these line profiles depend only on the line positions and temperature, and not on the (for example) quantum numbers. In fact, the main disadvantage of the histograms is that they lose any information on the upper and lower states, including the quantum numbers. This information is important when dealing with the pressure-dependent line profiles, which often show strong variation with quantum numbers, particularly $J$.
One can still assume, however, that the continuum is nearly featureless and thus not very sensitive to dependence of the line profiles on the quantum numbers of the upper or lower states.

\begin{figure}[!htbp]
  \includegraphics[width=0.48\textwidth]{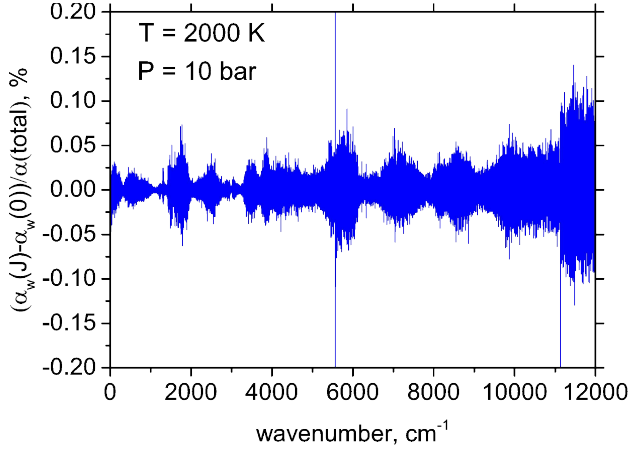}
\caption{\label{f:a0-J=0}
Relative error from using $J$-independent line broadening to describe methane continuum at high temperature ($T=2000$~K) and pressure ($P=10$~bar) as the difference between two cross sections ($J$-dependent $a_0$ model vs $J$-independent model) relative to the total cross sections. The wavenumber grid of $\Delta \tilde{\nu} = 0.1$~\cm\ is used.}
\end{figure}

In order to illustrate the applicability of this approximation in Figure~\ref{f:a0-J=0} we show the error of the methane continuum at $T=2000$~K and $P=10$~bar as the difference between two cross sections: (i) obtained using the $J$-dependent Voigt-profile model by \citet{jt684} and (ii) obtained using constant Voigt parameters, relative to the total methane cross sections at these values of $T$ and $P$. The error is within 0.05~\%\ for the most of the frequency range and not larger than 0.1~\%. Another artifact of the histogram method (apart from the limited profile description) is the error of the line position within a bin. Therefore the smaller the bin the better accuracy of the super-line list.

The important advantage of histograms is that they are very robust and efficient for computing cross sections due to a relatively small number of the super-lines defined by the density of the wavenumber grid, which is therefore much smaller (at least for methane) than the number of the original lines. For example, with the 0.01~\cm\ grid spacing, the size of a histogram at a given $T$ is only 1,200,000 grid points (super-lines) for our line list coverage ($<12,000$~\cm), which is much smaller than the original 34 billion lines. Even for a more sophisticated four-grids model suggested by \citet{jt631}  ($\Delta \tilde{\nu} = $ $10^{-5}$~\cm\ for 10--100~\cm, $10^{-4}$~\cm\ for 100--1,000~\cm, $0.001$~\cm\ for 1,000-10,000~\cm\ and 0.01~\cm\ for $>10,000$~\cm) we obtain 28,200,000 super-lines, which also should not be a problem for  line-by-line practical applications.
Since the long wavelength region is always more demanding in terms of the accuracy, such dynamic grids are more accurate. In the following we also propose another dynamic grid based on a constant resolving power, $R$.

In order to benchmark the super-line approach we have computed three sets of histograms for $T=2000$~K representing the continuum of methane (i.e. from the `weak' lines only) using the following grid models: histrogram~I is with a constant grid spacing of 0.01~\cm\ (1,200,000 points); histrogram~II consists of four sub-grids proposed by \citet{jt631} (28 million points); histrogram~III is constructed  to have a constant resolving power $R$ of 1,000,000 (7,090,081 points). The constant $R$-grid  can be defined to have variable grid spacings  as given by
$$
\frac{\tilde\nu}{\Delta \tilde{\nu}} = R.
$$
Thus the vavenumber grid  point $\tilde\nu_k$  ($k=0\ldots N(R)$) is given by:
\begin{equation}\label{e:super2}
\tilde\nu_k = \tilde\nu_A \, a^i,
\end{equation}
where $a = (R+1)/R$ and $\tilde\nu_A = \tilde\nu_0 $ is the left-most wavenumber grid point (\cm). The total number of bins,  $N(R)$ is given by:
\begin{equation}\label{e:super}
N(R) = \frac{\log\frac{\tilde\nu_B}{\tilde\nu_{A}}}{\log a},
\end{equation}
where $\tilde\nu_B = \tilde\nu_N$ is the right-most grid point
and $N(R)+1$ is the total number of the grid points.

\begin{figure}[!htbp]
  \includegraphics[width=0.48\textwidth]{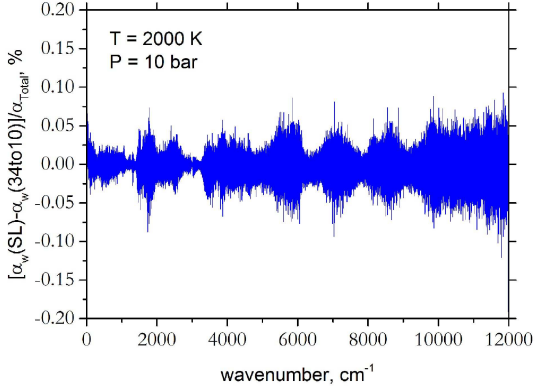}
\caption{\label{f:error-SL:R-1M}
Relative errors using the histogram-model $R=1,000,000$ to describe the methane continuum at $T=2000$~K  and $P=10$~bar as the difference with the 34to10 cross sections (Voigt-model) relative to the total 34to10  cross sections.  The wavenumber grid of $\Delta \tilde{\nu} = 0.01$~\cm\ is used.  }
\end{figure}

The histrograms I, II and III were used to generate the continuum cross sections of CH$_4$ at $P=10$~bar. Here we assumed the Voigt profile with constant parameters ($\gamma_0 = 0.051$~\cm, $n=0.44$, $T_0 = 298$~K and $P_0=$ 1 bar) and used the grid with $\Delta \tilde{\nu} = 0.01$~\cm. These cross sections were then compared to the corresponding continuum cross sections ($T=2000$~K, $P=10$~bar) computed line-by-line directly from the 34to10 line list. All
histogram models show very similar, almost identical deviations, well below 0.1~\%\ for the most of the range.  Figure~\ref{f:error-SL:R-1M} illustrates the relative  errors obtained for the $R=1,000,000$ histogram model.

\begin{figure}[!htbp]
  \includegraphics[width=0.5\textwidth]{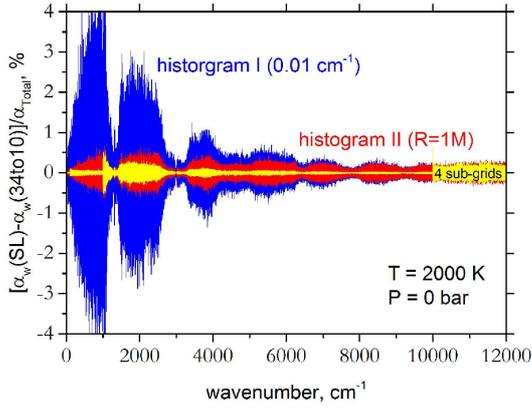}
\caption{\label{f:error-SL:3-grids:P=0bar}
Relative error from the histogram-model for three different grids to describe  the methane continuum at $T=2000$~K  and $P=0$~bar as the difference with the 34to10 cross sections (pure Doppler-model) relative to the total 34to10  cross sections at $P=0$~bar. The wavenumber grid of $\Delta \tilde{\nu} = 0.01$~\cm\ is used. }
\end{figure}

Now we turn to the case of the pure Doppler broadening ($P=0$~bar, $T=2000$~K, grid spacing $\Delta \tilde{\nu} = 0.01$~\cm), where the lines are sharper and narrower, such that the line width may become comparable or even smaller than the
grid spacing. Figure~\ref{f:error-SL:3-grids:P=0bar} illustrates the errors for the same three histogram models. The uniform histrogram~I of 0.01~\cm\ (1,200,000 points) exhibits the largest errors in the low frequency region, while the two adaptive grids show errors within about 4--5~\%. Clearly, $\Delta \tilde{\nu}$ = 0.01~\cm\ is too coarse for the super-line approach to describe the low frequency range in the the zero pressure case, therefore we recommend using grids with more points (lines) in the region below 1000~\cm.  For the denser histrograms II and III the error drops to $<$0.2--0.5~\%. The histogram~III (resolving power $R=1,000,000$) shows a more even error distribution.

Similar comparison for $T=300$~K showed even better agreement, with errors about an order of magnitude smaller than those found for $T=2000$~K. Using a coarser grid to simulate cross sections (e.g. $\Delta \tilde{\nu} = 0.1$~\cm) also drops the errors by an order of magnitude.

\begin{figure}[!htbp]
  \includegraphics[width=0.5\textwidth]{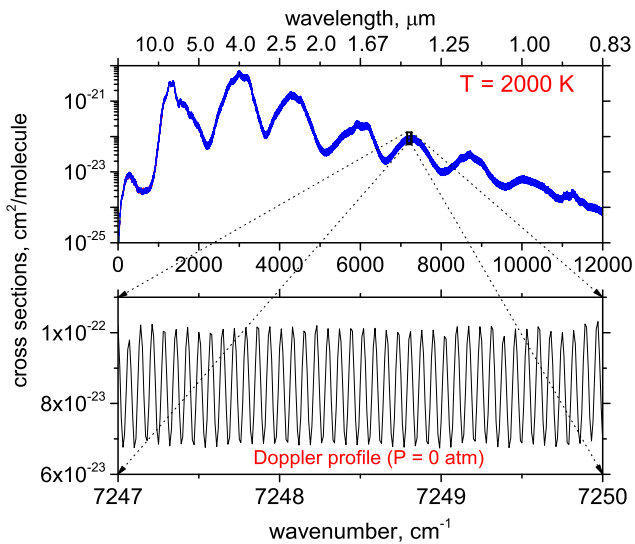}
\caption{\label{f:SL:coarse}
Example of a inappropriate use of the super-line approach when the grid is too coarse. The super-lines use a resolution of $R=100,000$, which is not sufficient due to the narrow Doppler profiles at zero pressure, $T=2000$~K. The cross section was computed on a $\Delta \tilde{\nu} = 0.01$~\cm\ grid.}
\end{figure}

For super-lines it is obviously important that the underlying grid spacing is not too large compared to the line width. This is illustrated in Fig.~\ref{f:SL:coarse}, which shows the $P=0$, $T=2000$~K continuum cross sections modelled using the $R=100,000$ histogram with the Doppler profile. It is clear that the Doppler line width is smaller than the separation between the super-lines, which leads to strong oscillations. In fact the same histogram performs well in case of much broader lines when modelling $P=10$~bar, Fig~\ref{f:a0-J=0}.

\begin{figure}[!htbp]
\includegraphics[width=0.48\textwidth]{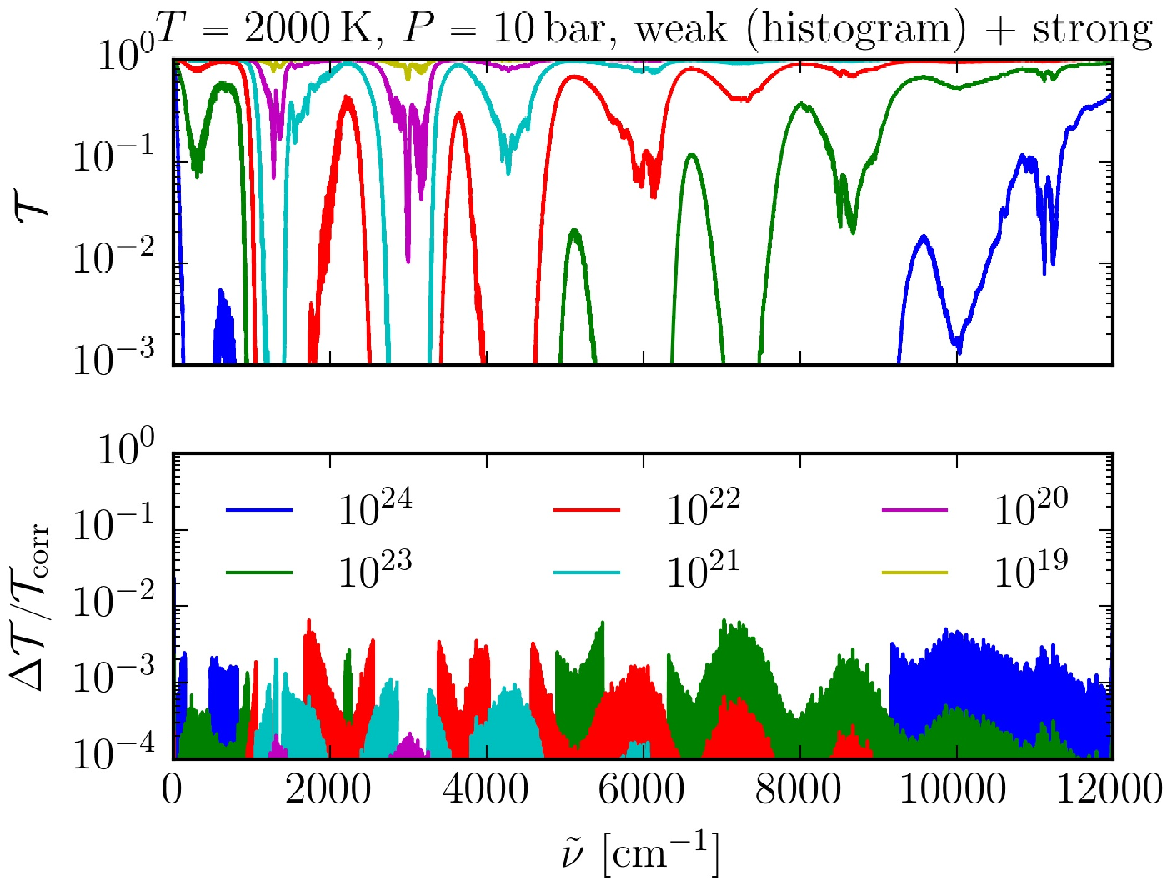}\\
\includegraphics[width=0.48\textwidth]{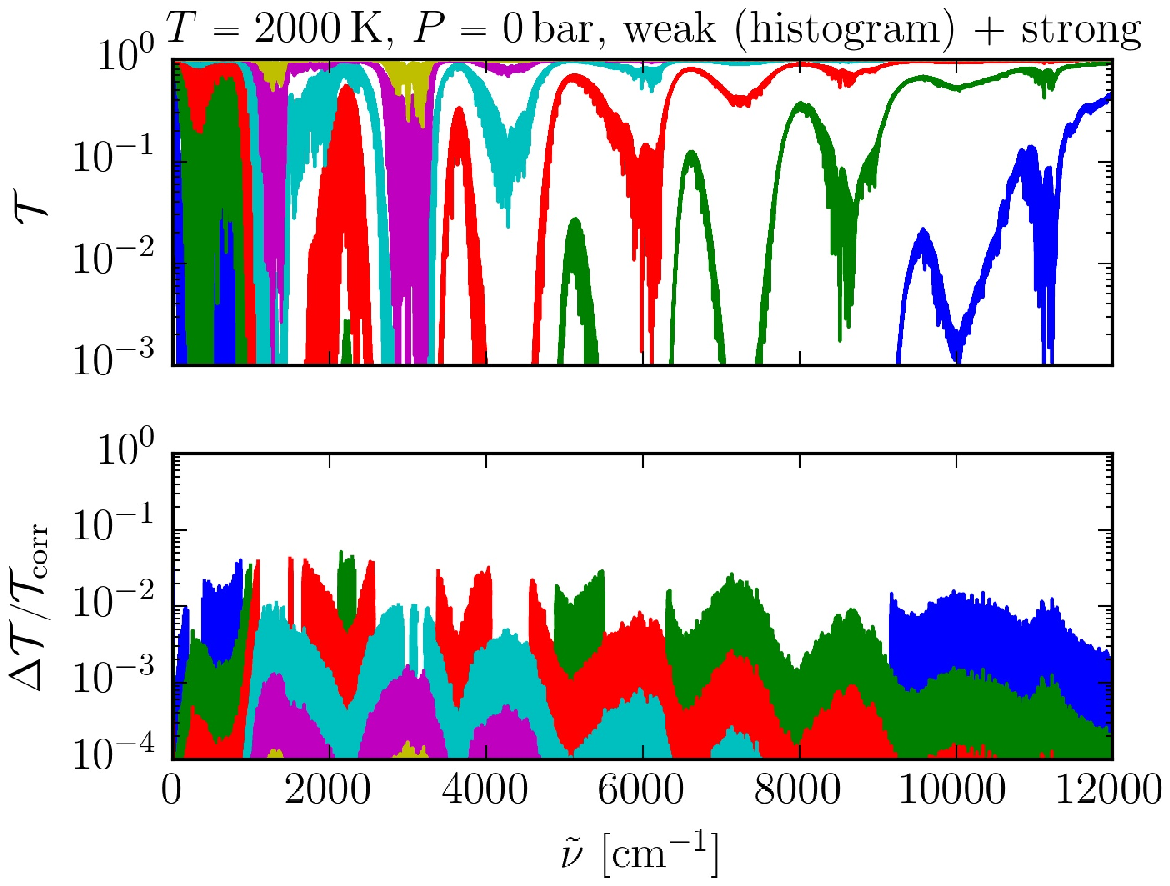}\\
\includegraphics[width=0.48\textwidth]{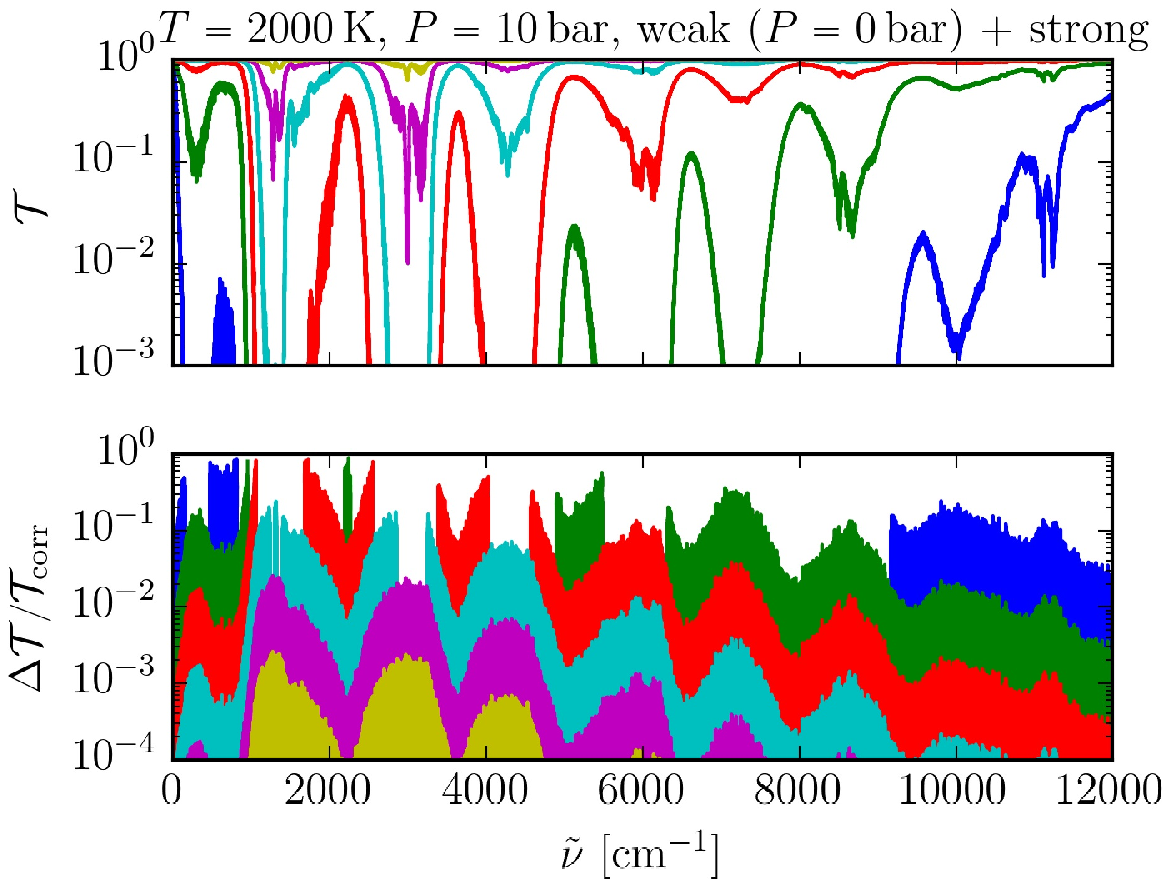}
  \caption{Transmissions computed using the Doppler (upper display) and histogram ($R=1,000,000$, middle and lower displays) continuum models with relative errors for the column amounts $10^{19}$, $10^{20}$, $10^{21}$, $10^{22}$, $10^{23}$, and $10^{24}$ molecule/cm$^2$ at $T=2000$~K, $P=0$ and $P=10\,$bar. The upper part of each display shows the total transmission obtained both from the strong weak lines at this temperature, while the lower part shows the relative error comparing to the direct line-by-line evaluation from the 34to10 line list using the \a0 Voigt model \protect\citep{jt684}. The error in regions with very small transmissions ($<10^{-4}$) are removed as the medium is optically thick.
 }
\label{f:transmit}
\end{figure}

In order to estimate the impact of the errors in the continuum models in actual atmospheric radiative transfer and retrieval calculations, we have calculated the transmission $\mathcal T$ and the relative error in the transmission $\Delta \mathcal T/\mathcal T_\text{corr}$ from the continuum models, where
\begin{equation}
\mathcal T = \exp \left[ -\alpha(\tilde \nu) u \right], \qquad
\frac{\Delta \mathcal T}{\mathcal T_\text{corr}} = \frac{\mathcal T - \mathcal T_\text{corr}}{\mathcal T_\text{corr}},
\end{equation}
and $\alpha(\tilde \nu)$ is the total cross section, $u$ is the column amount and $\mathcal T_\text{corr}$ is the correct transmission calculated from the direct line-by-line evaluation of the 34to10 line list using the \a0 Voigt model \protect\cite{jt684}. We show the transmissions and errors in Figure~\ref{f:transmit} obtained using both continuum models with column amounts ranging from $10^{19}$ to $10^{24}\,$molecule/cm$^2$ at $T=2000\,$K, and $P=0$ and $P=10\,$bar. The histogram model performs extremely well for the high pressure case (lower display) with the errors within $1\,$\%\ and significantly better than the Doppler-grid model (upper display). The errors in the histogram model at zero-pressure are higher due to the very narrow lines at small wavenumbers (middle display), but should be acceptable for most of the applications (within $5\,$\%). If higher accuracy is required, the histogram resolution should be increased.

\section{Hybrid line list and temperature dependent continuum cross sections}
\label{s:result}

Our partitioning of the total 34,170,582,862 lines in our new 34to10 line list leads to 16,776,857 strong and 34,153,806,005 weak lines. The latter were used to (i) generate temperature-dependent continuum cross sections (Doppler-broadened) and (ii) temperature-dependent histograms of super-lines
for the following set of temperatures: 296~K, 400~K, 500~K, 600~K, 700~K, 800~K, 900~K, 1000~K, 1100~K, 1200~K, 1300~K, 1400~K, 1500~K, 1600~K, 1700~K, 1800~K, 1900~K, and 2000~K.
A wavenumber grid with constant $R=1,000,000$ consisting of 7,090,081 points (super-lines)  was adopted for the total range of 0 -- 12000~\cm. The remaining 16,776,857 strong lines together with the .states file containing 8,194,057 energies form a line list in the standard ExoMol format \citep{jt631}. The super-lines are stored in the two-column format with the frequency wavenumbers (\cm) and absorption coefficients (cm$/$molecule), which is the same as the format used for the ExoMol cross sections \citep{jt631}.
Thus the histogram format does not require any information on the upper/lower states, temperature, partition function, or statistical weights, only the line profile specifications are needed.  The line broadening can only depend on the wavenumber.
The hybrid line list is given as supplementary material to this paper via the CDS database \url{http://cdsarc.u-strasbg.fr} and can also be found on the ExoMol website \url{www.exomol.com}. We also include the Voigt-model used in the simulations of cross sections.

\section{Conclusion}

We have extended our previous 10to10 methane line list to higher temperatures, the result of which is a new line list containing 34 billion transitions. Line lists of this size are impractical to work with as the calculation of cross sections becomes extremely computationally expensive. We have therefore explored the idea of partitioning this line list into a relatively small subset of strong lines which are retained and to be fully treated in any cross section calculation, and augmented by a temperature-dependent quasi-continuum which represents the contribution of the remaining lines. A key assumption is that assume the this quasi-continuum to be essentially featureless and not very sensitive to the variation of the pressure broadened line shape. The strong lines are selected such that they retain the flexibility required to describe the variation of the shape of the methane absorption with pressure.

Two $P$-independent models were tested to represent the continuum built from the weak lines, Doppler-broadened cross sections and super-lines. For the Doppler-broadened scheme, the assumption is that the methane continuum does not strongly depend on pressure and can be modelled using the pressure-independent line profiles (Doppler). The error of this approach on dense grids (0.01~\cm) ranges from within 8~\%\ for long wavelength down to within 3~\%\ above 1~$\mu$m. The coarser grid of 0.1~\cm\ gives the errors within 2~\%.

The super-lines approach is more flexible as it allows the continuum to depend on pressure. The variation with pressure, however, should not depend on the upper or lower states, only on the line position. For this  model we also introduced the dynamic grid representation with a constant resolution, with the grid spacing changing as a function of the wavenumber to keep $R = \tilde\nu/ \Delta \tilde\nu$ the same. Each grid in this histogram model point containing the total absorption within the $\Delta \tilde\nu$ bin is then used as a super-line. We find that the super-lines  built as histograms on an adaptive grid of a high resolution are more accurate for absorption modelling, and therefore was put forward as the ExoMol standard. The typical errors even for dense grids are within 1~\%. With our selected partitioning we retain 17 million strong lines for our strong line lists and computed a set of histograms containing 7,090,081 points (super-lines) for a set of 18 temperatures using a dynamic wavenumber grid with a resolution of $R=1,000,000$. The
strong lines are given as the ExoMol line list while the continuum histograms are presented using the ExoMol format developed for cross sections \citep{jt631}. We recommend to use this hybrid line list based on  the super-line approach for line-by-line atmospheric modeling of methane absorption. For low pressures and short-wavelengths, the resolution might need to be increased to higher than 1,000,000 due to the very narrow Doppler-broadened lines and their low density in this region.

The integrated errors of cross sections over an extended frequency range (significantly larger than than linewidth) are found to be vanishingly small if the line profiles used preserve the area (subject to numerical accuracy). That is, for optically thin atmospheres both continuum models will guarantee that exact answer for the integrated opacities. We have also shown that even in case of realistic, not optically thin media, the super-line approach leads to very small error transmission.


The data can be accessed via the CDS database \url{http://cdsarc.u-strasbg.fr} as well as at
\url{www.exomol.com}. These line lists can be downloaded from the CDS, via
ftp://cdsarc.u-strasbg.fr/pub/cats/J/A+A/, or
http://cdsarc.u-strasbg.fr/viz-bin/qcat?J/A+A/, or from www.exomol.com.



\section*{Acknowledgements}


This work was supported by the UK Science and Technology Research Council (STFC) No. ST/M001334/1, ERC Advanced Investigator Projects 267219 and 247060-PEPS, and the COST action MOLIM No. CM1405.  DSA acknowledges support from the NASA Astrobiology Program through the Nexus for Exoplanet System Science. This work made extensive use of the DiRAC@Darwin and DiRAC@COSMOS HPC  clusters.  DiRAC is the UK HPC facility for particle physics, astrophysics and cosmology which is supported by STFC and BIS. Some calculations for this paper were performed on the University of Exeter Supercomputer, a DiRAC Facility jointly funded by STFC, the Large Facilities Capital Fund of BIS, and the University of Exeter.


\bibliographystyle{aa}

\begin{thebibliography}{31}
\expandafter\ifx\csname natexlab\endcsname\relax\def\natexlab#1{#1}\fi

\bibitem[{Al-Refaie {et~al.}(2015{\natexlab{a}})Al-Refaie, Ovsyannikov,
  Polyansky, Yurchenko, \& Tennyson}]{jt620}
Al-Refaie, A.~F., Ovsyannikov, R.~I., Polyansky, O.~L., Yurchenko, S.~N., \&
  Tennyson, J. 2015{\natexlab{a}}, J. Mol. Spectrosc., 318, 84

\bibitem[{Al-Refaie {et~al.}(2015{\natexlab{b}})Al-Refaie, Yurchenko,
  Yachmenev, \& Tennyson}]{jt597}
Al-Refaie, A.~F., Yurchenko, S.~N., Yachmenev, A., \& Tennyson, J.
  2015{\natexlab{b}}, MNRAS, 448, 1704

\bibitem[{Amundsen {et~al.}(2014)Amundsen, Baraffe, Tremblin, Manners,
  Wolfgang, Mayne, \& Acreman}]{14AmBaTr.broad}
Amundsen, D.~S., Baraffe, I., Tremblin, P., {et~al.} 2014, A\&A, 564, A59

\bibitem[{Amundsen {et~al.}({2016})Amundsen, Mayne, Baraffe, Manners, Tremblin,
  Drummond, Smith, Acreman, \& Homeier}]{16AmMaBa}
Amundsen, D.~S., Mayne, N.~J., Baraffe, I., {et~al.} {2016}, A\&A, {595}, A36

\bibitem[{Bailey \& Kedziora-Chudczer({2012})}]{12BaKe}
Bailey, J. \& Kedziora-Chudczer, L. {2012}, MNRAS, {419}, 1913

\bibitem[{Barton {et~al.}(2017)Barton, Hill, Czurylo, Li, Hyslop, Yurchenko, \&
  Tennyson}]{jt684}
Barton, E.~J., Hill, C., Czurylo, M., {et~al.} 2017, J. Quant. Spectrosc.
  Radiat. Transf.

\bibitem[{Canty {et~al.}(2015)Canty, Lucas, Yurchenko, Tennyson, Leggett,
  Tinney, Jones, Burningham, Pinfield, \& Smart}]{15CaLuYu.dwarfs}
Canty, J.~I., Lucas, P.~W., Yurchenko, S.~N., {et~al.} 2015, MNRAS, 450, 454

\bibitem[{{Drummond} {et~al.}(2016){Drummond}, {Tremblin}, {Baraffe},
  {Amundsen}, {Mayne}, {Venot}, \& {Goyal}}]{16DrTrBa}
{Drummond}, B., {Tremblin}, P., {Baraffe}, I., {et~al.} 2016, A\&A, 594, A69

\bibitem[{Hargreaves {et~al.}(2015)Hargreaves, Bernath, Bailey, \&
  Dulick}]{15HaBeBa.CH4}
Hargreaves, R.~J., Bernath, P.~F., Bailey, J., \& Dulick, M. 2015, ApJ, 813, 12

\bibitem[{Hill {et~al.}(2013)Hill, Yurchenko, \& Tennyson}]{jt542}
Hill, C., Yurchenko, S.~N., \& Tennyson, J. 2013, Icarus, 226, 1673

\bibitem[{Irwin {et~al.}(2008)Irwin, Teanby, {de Kok}, Fletcher, Howett, Tsang,
  Wilson, Calcutt, Nixon, \& Parrish}]{NEMESIS}
Irwin, P. G.~J., Teanby, N.~A., {de Kok}, R., {et~al.} 2008, J. Quant.
  Spectrosc. Radiat. Transf., 109, 1136

\bibitem[{{Malik} {et~al.}(2017){Malik}, {Grosheintz}, {Mendon{\c c}a},
  {Grimm}, {Lavie}, {Kitzmann}, {Tsai}, {Burrows}, {Kreidberg}, {Bedell},
  {Bean}, {Stevenson}, \& {Heng}}]{HELIOS}
{Malik}, M., {Grosheintz}, L., {Mendon{\c c}a}, J.~M., {et~al.} 2017, ApJ, 153,
  56

\bibitem[{Nikitin {et~al.}(2017)Nikitin, Rey, \& Tyuterev}]{17NiReTy}
Nikitin, A.~V., Rey, M., \& Tyuterev, V.~G. 2017, J. Quant. Spectrosc. Radiat.
  Transf.,

\bibitem[{Rey {et~al.}(2016)Rey, Nikitin, Babikov, \& Tyuterev}]{TheoReTS}
Rey, M., Nikitin, A.~V., Babikov, Y.~L., \& Tyuterev, V.~G. 2016, J. Mol.
  Spectrosc., 327, 138

\bibitem[{Rey {et~al.}({2014})Rey, Nikitin, \& Tyuterev}]{14ReNiTy.CH4}
Rey, M., Nikitin, A.~V., \& Tyuterev, V.~G. {2014}, ApJ, {789}, 2

\bibitem[{Rothman {et~al.}(2013)Rothman, Gordon, Babikov, Barbe, Benner,
  Bernath, Birk, Bizzocchi, Boudon, Brown, Campargue, Chance, Cohen, Coudert,
  Devi, Drouin, Fayt, Flaud, Gamache, Harrison, Hartmann, Hill, Hodges,
  Jacquemart, Jolly, Lamouroux, {Le Roy}, Li, Long, Lyulin, Mackie, Massie,
  Mikhailenko, M{\"u}ller, Naumenko, Nikitin, Orphal, Perevalov, Perrin,
  Polovtseva, Richard, Smith, Starikova, Sung, Tashkun, Tennyson, Toon,
  Tyuterev, \& Wagner}]{HITRAN2012}
Rothman, L.~S., Gordon, I.~E., Babikov, Y., {et~al.} 2013, J. Quant. Spectrosc.
  Radiat. Transf., 130, 4

\bibitem[{Sousa-Silva {et~al.}(2015)Sousa-Silva, Al-Refaie, Tennyson, \&
  Yurchenko}]{jt592}
Sousa-Silva, C., Al-Refaie, A.~F., Tennyson, J., \& Yurchenko, S.~N. 2015,
  MNRAS, 446, 2337

\bibitem[{Tennyson {et~al.}(2014)Tennyson, Bernath, Campargue, Cs\'asz\'ar,
  Daumont, Gamache, Hodges, Lisak, Naumenko, Rothman, Tran, Zobov, Buldyreva,
  Boone, {De Vizia}, Gianfrani, Hartmann, McPheat, Murray, Ngo, Polyansky, \&
  Weidmann}]{jt584}
Tennyson, J., Bernath, P.~F., Campargue, A., {et~al.} 2014, Pure Appl. Chem.,
  86, 1931

\bibitem[{Tennyson \& Yurchenko(2012)}]{jt528}
Tennyson, J. \& Yurchenko, S.~N. 2012, MNRAS, 425, 21

\bibitem[{Tennyson \& Yurchenko(2017)}]{jt693}
Tennyson, J. \& Yurchenko, S.~N. 2017, Mol. Astrophys., 8, 1

\bibitem[{Tennyson {et~al.}(2016)Tennyson, Yurchenko, Al-Refaie, Barton, Chubb,
  Coles, Diamantopoulou, Gorman, Hill, Lam, Lodi, McKemmish, Na, Owens,
  Polyansky, Rivlin, Sousa-Silva, Underwood, Yachmenev, \& Zak}]{jt631}
Tennyson, J., Yurchenko, S.~N., Al-Refaie, A.~F., {et~al.} 2016, J. Mol.
  Spectrosc., 327, 73

\bibitem[{Tremblin {et~al.}(2016)Tremblin, Amundsen, Chabrier, Baraffe,
  Drummond, Hinkley, Mourier, \& Venot}]{16TrAmCh}
Tremblin, P., Amundsen, D.~S., Chabrier, G., {et~al.} 2016, ApJL, 817, L19

\bibitem[{{Tremblin} {et~al.}(2015){Tremblin}, {Amundsen}, {Mourier},
  {Baraffe}, {Chabrier}, {Drummond}, {Homeier}, \& {Venot}}]{15TrAmMo}
{Tremblin}, P., {Amundsen}, D.~S., {Mourier}, P., {et~al.} 2015, ApJL, 804, L17

\bibitem[{Underwood {et~al.}(2016)Underwood, Tennyson, Yurchenko, Clausen, \&
  Fateev}]{jt641}
Underwood, D.~S., Tennyson, J., Yurchenko, S.~N., Clausen, S., \& Fateev, A.
  2016, MNRAS, 462, 4300

\bibitem[{Waldmann {et~al.}(2015{\natexlab{a}})Waldmann, Rocchetto, Tinetti,
  Barton, Yurchenko, \& Tennyson}]{jt611}
Waldmann, I.~P., Rocchetto, M., Tinetti, G., {et~al.} 2015{\natexlab{a}}, ApJ,
  813, 13

\bibitem[{Waldmann {et~al.}(2015{\natexlab{b}})Waldmann, Tinetti, Barton,
  Yurchenko, \& Tennyson}]{jt593}
Waldmann, I.~P., Tinetti, G., Barton, E.~J., Yurchenko, S.~N., \& Tennyson, J.
  2015{\natexlab{b}}, ApJ, 802, 107

\bibitem[{Yurchenko {et~al.}(2017)Yurchenko, Al-Refaie, \& Tennyson}]{ExoCross}
Yurchenko, S.~N., Al-Refaie, A.~F., \& Tennyson, J. 2017, {ExoCross: a set of
  tools to work with molecular line lists}

\bibitem[{Yurchenko \& Tennyson(2014)}]{jt564}
Yurchenko, S.~N. \& Tennyson, J. 2014, MNRAS, 440, 1649

\bibitem[{Yurchenko {et~al.}(2014)Yurchenko, Tennyson, Bailey, Hollis, \&
  Tinetti}]{jt572}
Yurchenko, S.~N., Tennyson, J., Bailey, J., Hollis, M. D.~J., \& Tinetti, G.
  2014, Proc. Nat. Acad. Sci., 111, 9379

\bibitem[{Yurchenko {et~al.}(2013)Yurchenko, Tennyson, Barber, \&
  Thiel}]{jt555}
Yurchenko, S.~N., Tennyson, J., Barber, R.~J., \& Thiel, W. 2013, J. Mol.
  Spectrosc., 291, 69

\bibitem[{Yurchenko {et~al.}(2007)Yurchenko, Thiel, \& Jensen}]{TROVE}
Yurchenko, S.~N., Thiel, W., \& Jensen, P. 2007, J. Mol. Spectrosc., 245, 126

\end{thebibliography}

\end{document}